\documentclass[aps,prd,onecolumn,superscriptaddress,showpacs,floatfix,nofootinbib]{revtex4}%% For Scientific Linux
\usepackage{psfig}
\usepackage{comment}
\usepackage{graphicx}
\usepackage[dvips]{color}
\usepackage{bm}

\begin{document}

\vspace{2cm}
\begin{flushright}
%\preprint
%{CDF/PUB/EXOTIC/PUBLIC/10921 V2 \\
%\preprint{
%FERMILAB-CONF-12-723-E}
  \vspace{0.5in}
\end{flushright}

%%%%%\markboth{Weiming Yao}
%%%%%{SM Higgs Searches at the Tevatron}

%%%%%\catchline{}{}{}{}{}
%
%%%%%%%%%%%%%%%%%%%%%%%%%%%%%%%%%%%%%%%%%%%%%%%%%%%%%%%%%%%%%%%%%%%%

\title{Studies of measuring Higgs self-coupling with 
$HH\rightarrow b\bar b \gamma\gamma$ at the future hadron colliders}

\author{Weiming Yao\footnote{Email contact: wmyao@lbl.gov}}

\affiliation{LBNL, Berkeley CA 94720, USA}

%%%%%%%%%%%%%\maketitle

%%%%%%%%\begin{history}
%%%%%%%\received{4 Sept. 2012}
%%%%%%%\revised{8 Sept. 2012}
%%%%%%%\end{history}

\begin{abstract}
We present a feasibility study of observing $HH\rightarrow b\bar b\gamma\gamma$
at the future hadron colliders with $\sqrt{s}=$14, 33, and 100 TeV. 
The measured cross section then can be used to 
constrain the Higgs self-coupling directly in the standard model. 
Any deviation could be a sign of new physics. The signal and background events 
are estimated using Delphes 3.0.10 fast Monte Carlo simulation based on the 
ATLAS detector capabilities. With 3 ab$^{-1}$ data, it would be possible to 
measure the Higgs self-coupling with a 50\%, 20\%, and 8\% statistical accuracy 
by observing $HH\rightarrow b\bar b\gamma\gamma$ at 
$\sqrt{s}=$14, 33, and 100 TeV colliders, respectively.  
 
%\smallskip
%\noindent \textbf{Keywords: Tevatron; CDF; D0; Higgs boson Search}

%%%%%%%%%%%\keywords{Tevatron; CDF; D0; Higgs boson Search}
\end{abstract}

%%%%%%%%%\ccode{PACS numbers: 13.85.Rm, 14.80.Bn}
%\pacs{PACS numbers: 13.85.Rm, 14.80.Bn}

\maketitle

\section{Introduction}	
\begin{figure}[htpb]
\centerline{\psfig{file=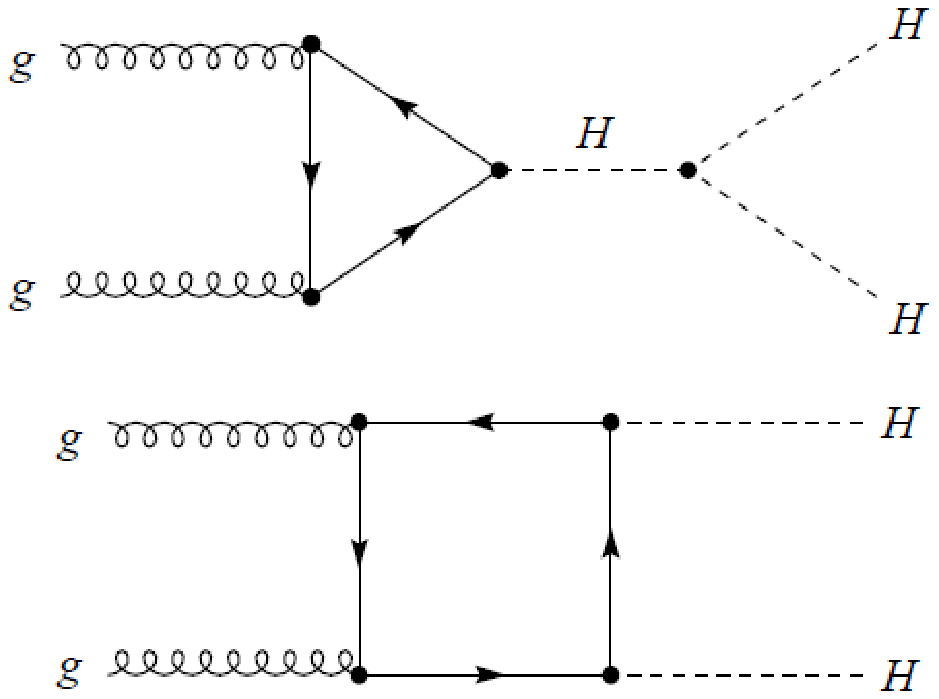,width=7.4cm} %6.7 cm
\psfig{file=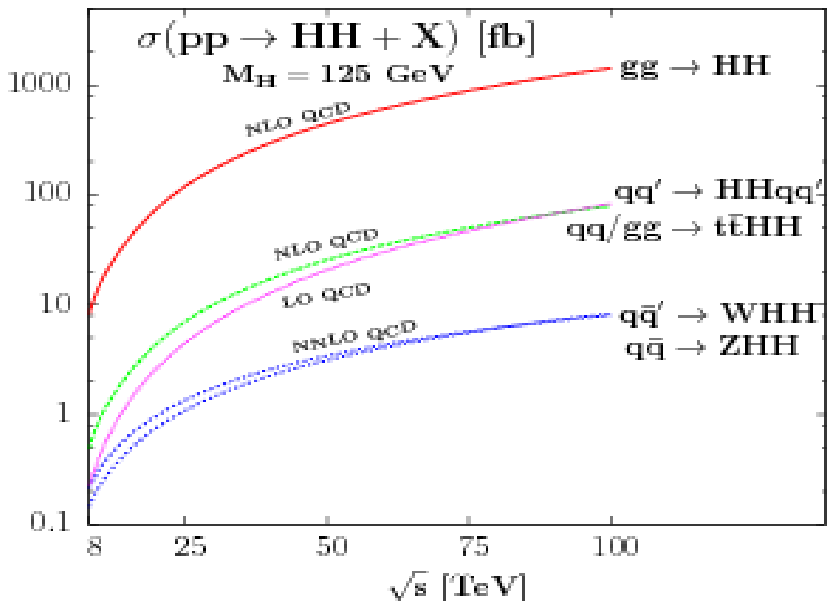,width=7.4cm}} %4.7cm
\vspace*{8pt}
\caption{The Feynman diagrams for the $gg\rightarrow HH$ processes (left) and
the corresponding production cross section as function of the collider energy
$\sqrt{s}$ (right).}
\label{fig:diag}
\end{figure}

The ATLAS and CMS have recently discovered a new boson with a mass near 125 
GeV/c$^2$~\cite{discovery}, which is a giant leap for science. The updated 
results are consistent with the expectation of a Higgs boson~\cite{higgs}, 
the missing cornerstone of particle physics. 
In order to directly test whether 
the Higgs mechanism is responsible for the electroweak symmetry breaking, we need to 
determine the Higgs self-coupling constants $\lambda_{HHH}$ directly from data
by observing the double Higgs production process $gg\rightarrow 
H\rightarrow HH$. In the standard model (SM), the Higgs self-coupling 
$\lambda_{HHH}$ is equal to $3M_H^2/v$ where $v=246$ GeV and $M_H$ is 
the measured Higgs boson mass. An accurate test of this relation may reveal 
the extended nature of the Higgs sector, which can be achieved by observing a 
significant deviation from the SM prediction above~\cite{ilc,baur04,zurita,baglio}.   
Recent studies indicate that observing $gg\rightarrow H\rightarrow H H$ is 
challenging at the high luminosity run of LHC (HL-LHC) with an integrated 
luminosity of 3000 fb$^{-1}$~\cite{atlashh}, due to destructive interference between $HHH$ and
$gg\rightarrow HH$ processes that are shown in Fig.~\ref{fig:diag} (left). 

%\begin{figure}[htpb]
%\centerline{\psfig{file=D0_tautau_limits.eps,width=4.7cm}
%\psfig{file=cdf_newlimitsBless_tth_limit.eps,width=4.7cm}
%\psfig{file=tevgamgambayeslimits19jun2012.eps,width=4.7cm}}
%\vspace*{8pt}
%\caption{The limits obtained from the searches on $H\rightarrow \tau^+\tau^-$ 
%from D0 (left), $t\bar t H$ from CDF (middle), and $H\rightarrow \gamma\gamma$
%from the Tevatron (left), respectively.\label{fig:secondsearch}}
%\end{figure}

In this study we study how feasible it is to measure Higgs self-coupling at the LHC 
and the future higher energy colliders(VLHC). We will focus 
on $HH\rightarrow b\bar b\gamma\gamma$ as the baseline. The advantage of 
measuring the Higgs self-coupling at the higher energy colliders is their large 
production cross section rate as shown in Fig.~\ref{fig:diag} (right), which increases from 
34 fb to 1418 fb when increasing the center mass of energy from 14 to 100 TeV~\cite{baglio}.
Recent studies indicate that the resummation effects will further increase the NLO cross section
by 20\%-30\% and reduce the scale uncertainties~\cite{NNLO}, which would 
improve the chance of measuring the Higgs self-coupling at HL-LHC.          

\section{Simulation setup}
\begin{figure}[htpb]
\centerline{\psfig{file=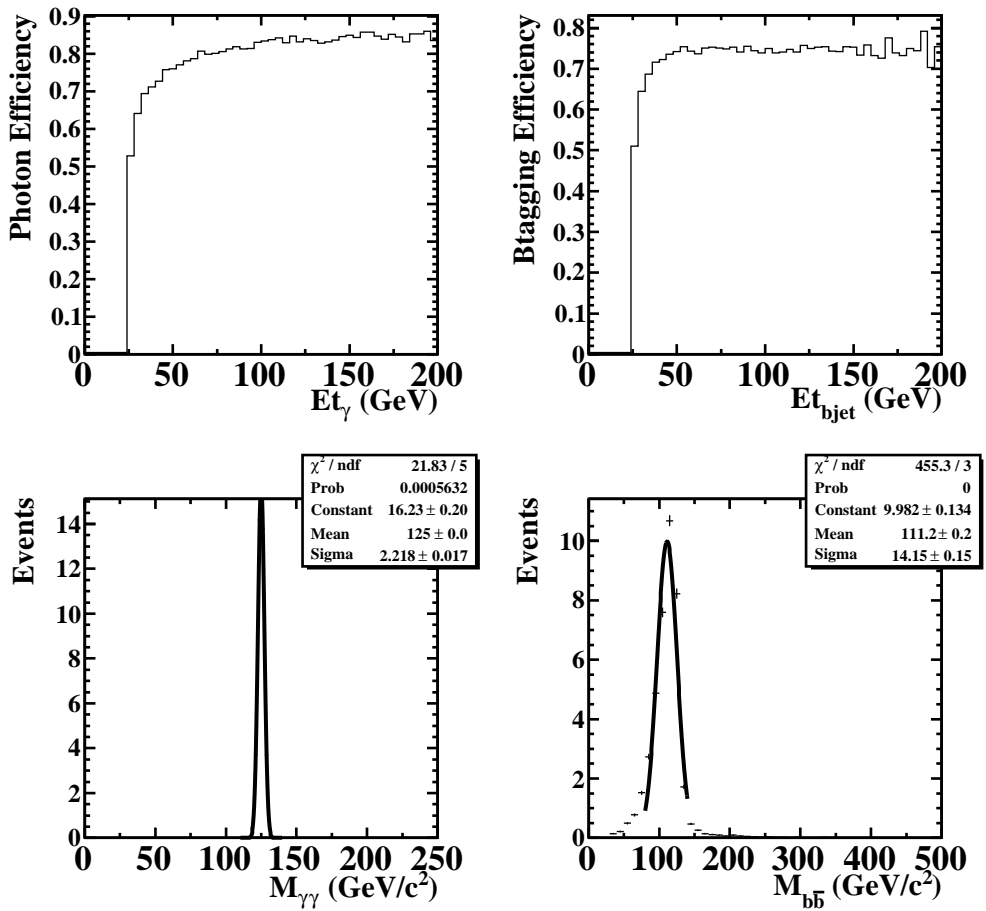,width=13.4cm}} %4.7cm
\vspace*{8pt}
\caption{The identification efficiency of photon as a function of photon $Et$ 
(upper left); the efficiency for $b$-jets (upper right); the invariant mass 
of $H\rightarrow \gamma\gamma$ (bottom left); and the invariant mass of 
$H\rightarrow b\bar b$ (bottom right).}
\label{fig:eff}
\end{figure}

Following the community summer studies 2013 (CSS 2013) guidelines, 
we use Delphes~\cite{delphes} V3.0.10 to simulate the ATLAS detector responses. More 
specifically, the photon energy is smeared according to the electromagnetic calorimeter 
(Ecal) responses of $\sigma_{E_T}/E_T = 0.20/\sqrt{E_T}\oplus 0.17\% $. The jet is clustered using the 
anti-$K_T$ algorithm with a radius of 0.5. The $b$-tagging operation point is 
chosen to have 75\% of efficiency and 1\% of mistags. In Fig.~\ref{fig:eff}, 
we show the identification efficiencies for photons and $b$-jets as a 
function of $Pt$ as well as the invariant mass distributions for $H\rightarrow \gamma\gamma$ and 
$H\rightarrow b\bar b$, respectively. The photon identification efficiency is about 80\% for 
photon with $E_T>25$ GeV and $|\eta|<2.5$.

The $gg\rightarrow HH$ signal is generated using HPAIR + PYTHIA6.2 
package~\cite{hpair}. All tree-level background processes up to 1 or 2 
partons are generated using Madgraph 5~\cite{madgraph} + PYTHIA8.0~\cite{pythia8.0} with MLM 
matching~\cite{mlm} to avoid double counting in certain regions of phase space. 
The production cross section of signal and background is evaluated 
using the CTEQ6L parton distribution functions~\cite{cteq6l} with the corresponding 
value of $\alpha_s$ at the investigated order in perturbative QCD.  
The signal and background processes for their cross section times branching ratio 
and the number of generated events are summarized in Table~\ref{tab:table1} for 
the colliders with $\sqrt{s}=$14, 33, and 100 TeV.

\begin{table}
\caption{The signal and background processes of production cross section times branching ratio and
the number of generated events for the colliders with $\sqrt{s}=$14, 33, and 100 TeV,}
\label{tab:table1}
\vspace{.5cm}
\begin{tabular}{|c|c|cc|cc|cc|}
\hline \hline
Samples&Gen. cuts & \multicolumn{2}{|c|}{ HL-LHC } & \multicolumn{2}{|c|}{TeV33} & \multicolumn{2}{|c|}{TeV100} \\ \hline
 &    & $\sigma \cdot B$ (fb) & Eevent &   $\sigma \cdot{B}$ (fb) & Events & $\sigma \cdot{B}$ (fb) & Events \\
\hline
$H(b\bar b) H(\gamma\gamma)$  &  & 0.0892 & 80000 & 0.545 & 80000 & 3.73 & 80000 \\ \hline 
$b\bar b \gamma\gamma$    & $Et_{j,b,\gamma}>20,20,25$ & 294 & 1033875 & 1085 & 952811 &5037 & 763962\\
$Z(b\bar b) H(\gamma\gamma)$  & $Et_{j,b,\gamma}>20,0,20$ & 0.109 & 97168 & 0.278 & 82088 &0.876 & 68585 \\
$b\bar b H(\gamma\gamma)$   & $Et_{j,b,\gamma}>20,0,20$ & 2.23 & 120617 & 9.843 & 110663 &50.49 & 99611 \\
$t\bar t H(\gamma\gamma)$ & $Et_{j,b,\gamma}>20,0,20$ & 0.68 & 83491 &4.76 & 71790 &37.26 & 63904 \\
\hline
\hline

\hline \hline
\end{tabular}
\end{table}

\section{Event kinematics and selections}
\begin{figure}[htpb]
\centerline{\psfig{file=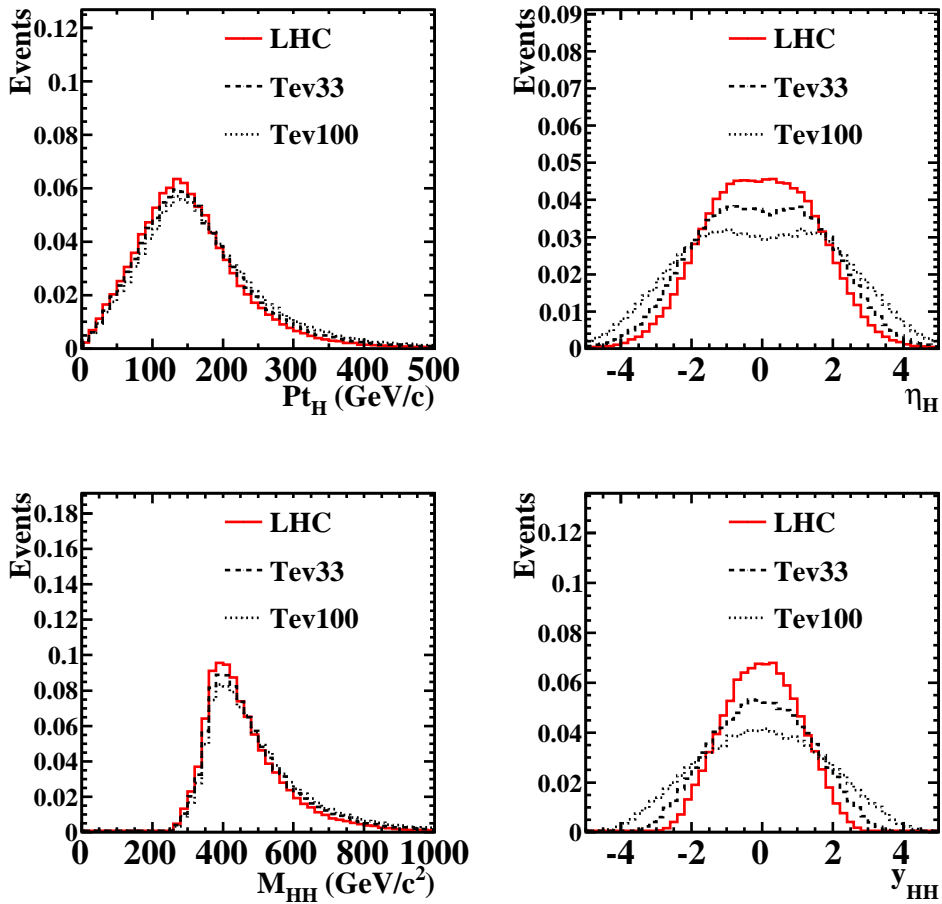,width=13.4cm}} %4.7cm
\vspace*{8pt}
\caption{The normalized distributions for $Pt_H$, $\eta_H$, $M_{HH}$, and 
$y_{HH}$ at the colliders with $\sqrt{s}=$14, 33, 100 TeV, respectively.}
\label{fig:fighh}
\end{figure}

The characteristic distributions of the gluon fusion process 
$gg\rightarrow H\rightarrow H H $ are compared for several observables at the 
hadron colliders with $\sqrt{s}=$14, 33, and 100 TeV. 
In Fig.~\ref{fig:fighh}, we show for the Higgs pairs 
the normalized distributions of the transverse momentum $Pt_H$, the pseudorapidity 
$\eta_H$, the invariant mass $M_{HH}$ ,and the rapidity $y_{HH}$.
They seem quite similar between the colliders 
so we use the common set of event selections to separate the signal from 
the backgrounds. The photons (npho) are required to be isolated and have $Et>25$ GeV 
and $|\eta|<2.5$. The jets (njet) are required to have $Et>25$ GeV and 
$|\eta|<2.5$. The $b$-jet candidate is a jet that has a $b$-tag. 
We select two $b$-jets and two photons in the final states to be consistent with the
signature of $gg\rightarrow HH \rightarrow b\bar b\gamma\gamma$ where each of the 
$b$-jets and photons is required to $Et>35$ GeV.
The invariant mass of two photons is then required to be consistent within 
5 GeV/c$^2$ of $M_H=125$ GeV/c$^2$ while the invariant mass of two $b$-jets is 
required to be between 85 and 135 GeV/c$^2$. In order to reject $t\bar t$ 
events, we also identify the number of isolated electrons and muons (nleps) 
with $Et(Pt)>25$ and $|\eta|<2.5$. If there is missing $Et>50$ GeV, we count 
nmet=1, otherwise nmet=0. 

\begin{figure}[htpb]
\centerline{\psfig{file=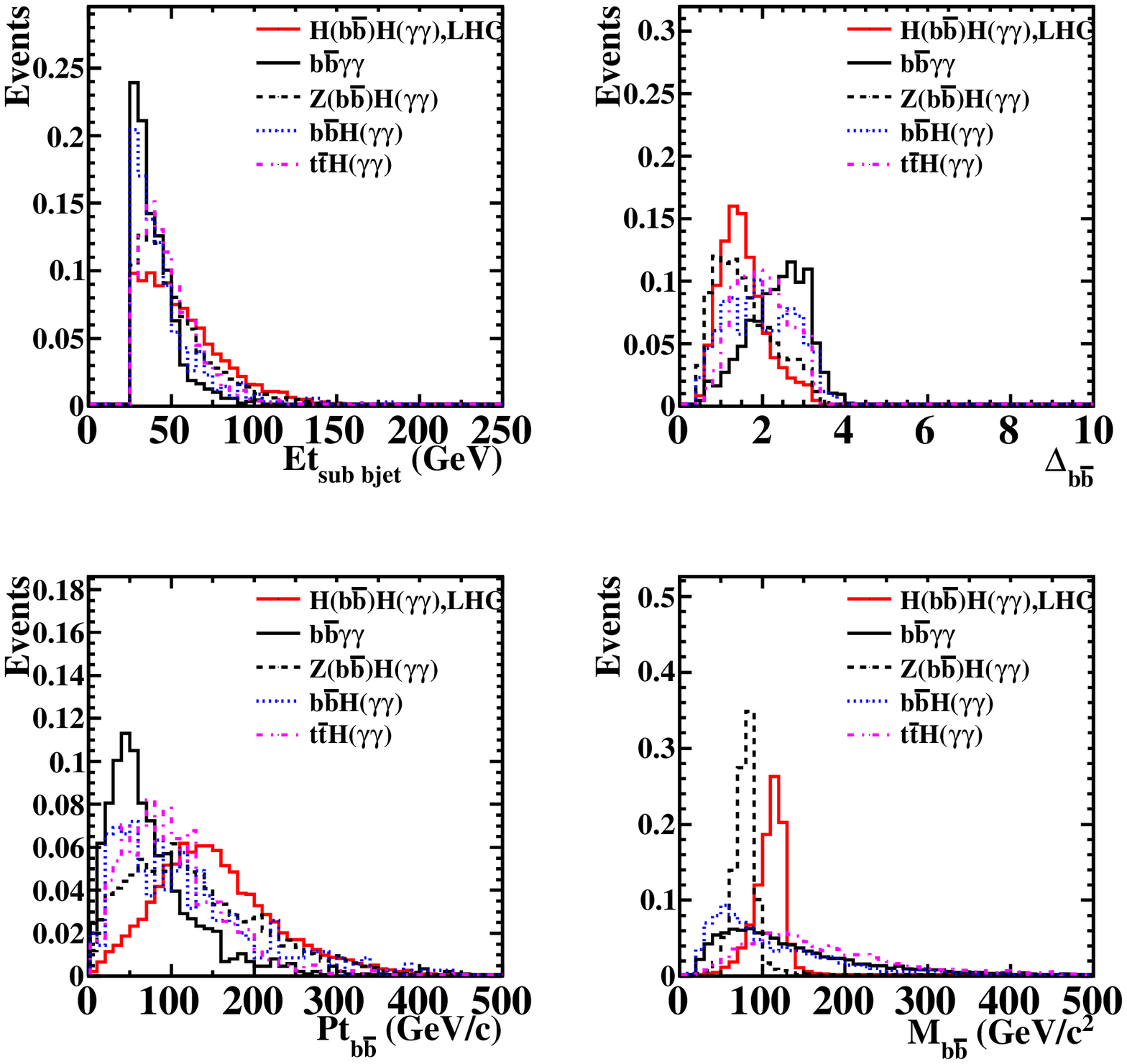,width=13.4cm}} %4.7cm
\vspace*{8pt}
\caption{The normalized distributions for the sub-leading $b$-jet $Et$, 
$\Delta R_{b\bar b}$, $Pt_{b\bar b}$, and $M_{b\bar b}$ from the signal and 
various background processes.}
\label{fig:figbb}
\end{figure}

\begin{figure}[htpb]
\centerline{\psfig{file=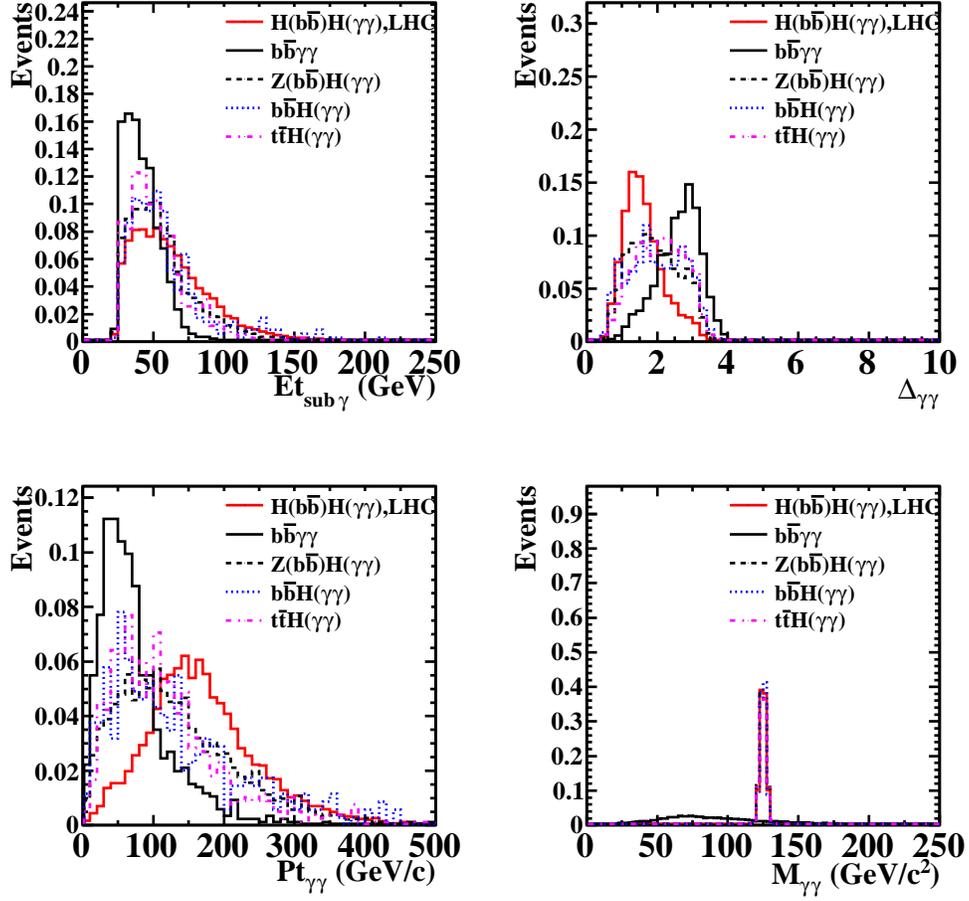,width=13.4cm}} %4.7cm
\vspace*{8pt}
\caption{The normalized distributions for the sub-leading photon $Et$,
$\Delta R_{\gamma\gamma}$, $Pt_{\gamma\gamma}$, and $M_{\gamma\gamma}$ from
the signal and various background processes.}
\label{fig:figgg}
\end{figure}

\begin{figure}[htpb]
\centerline{\psfig{file=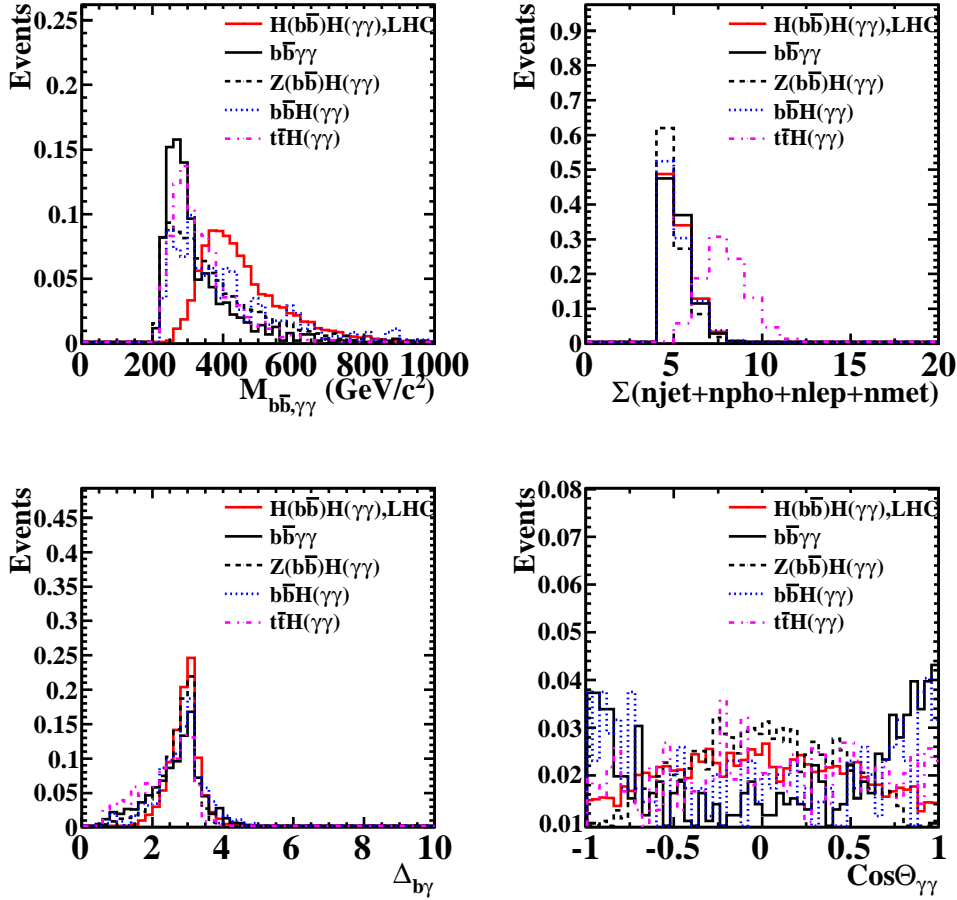,width=13.4cm}} %4.7cm
\vspace*{8pt}
\caption{The normalized distributions for $M_{b\bar b, \gamma\gamma}$, 
$\Sigma(njet+npho+nlep+nmet)$, $\Delta_{b\gamma,min}$, and $Cos\theta_{\gamma\gamma}$
from the signal and various background processes.}
\label{fig:figdh}
\end{figure}

For $H\rightarrow b\bar b$, we compare the kinematic 
distributions between the signal and backgrounds for the sub-leading 
$Pt_b$, the $\Delta R$ separation, the $Pt_{b\bar b}$, and the invariant mass of 
$M_{b\bar b}$ as shown in Fig.~\ref{fig:figbb}. For $H\rightarrow \gamma\gamma$,
the photon kinematic distributions are shown in Fig.~\ref{fig:figgg} for the 
sub-leading $Pt_{\gamma}$, the $\Delta R$ separation, the $Pt_{\gamma\gamma}$, and
the invariant mass of $M_{\gamma\gamma}$. We also compare the 
kinematic distributions of the pair of Higgs between the signal and backgrounds
for the invariant mass of $M_{b\bar b\gamma\gamma}$, 
$\Sigma(njet+npho+nlep+nmet)$, the minimum $\Delta R$ between the photons and 
the $b$-jets, and the $cos{\theta_{\gamma\gamma}}$, as shown in 
Fig.~\ref{fig:figdh}.

Based on these distributions, we further apply the following cuts to optimize
the sensitivity: 
\begin{itemize} 
  \item $\Delta R_{\gamma\gamma} <2.5$ and $\Delta R_{b\bar b}<2.0$
  \item $|\eta_{\gamma\gamma}|<2.0$ and $|\eta_{b\bar b}|<2.0$  
  \item $Pt_{\gamma\gamma}>100$ and $Pt_{b\bar b}>100$ GeV
  \item $M_{b\bar b\gamma\gamma}>300$ GeV/c$^2$
  \item $|Cos\theta_H |<0.8$, the Higgs decay angle in the rest frame of HH.
  \item $\Sigma(njets+nphos+nleps+nmet)<7$  
\end{itemize} 

\section{Preliminary Results} 
After applying the event selection described above, the remaining number of 
signal and background events are summarized in Table~\ref{tab:tabsummary} 
for an integrated luminosity of 3000 fb$^{-1}$. The background seems dominated 
by the QCD production of $b\bar b \gamma\gamma$, which can be further reduced 
using a multivariant analysis technique once a realistic simulation is 
available.

\begin{table}
\caption{The signal and background processes of $\sigma\times Br$, acceptance, and the expected events with 3000 fb$^{-1}$ data for the colliders with $\sqrt{s}=$14, 33, and 100 TeV.}
\label{tab:tabsummary}
\vspace{.5cm}
\begin{tabular}{|c|ccc|ccc|ccc|}
\hline \hline
Samples & \multicolumn{3}{|c|}{ HL-LHC (3 ab$^{-1}$) } & \multicolumn{3}{|c|}{TeV33 (3 ab$^{-1}$)} & \multicolumn{3}{|c|}{TeV100 (3 ab$^{-1}$)} \\ \hline
   & $\sigma \cdot Br$ & Acc. & Expect & $\sigma \cdot Br$ &  Acc. &  Expect &$\sigma \cdot Br$ &  Acc. & Expect \\  \hline
          & (fb) & (\%) & Evnts & (fb) & (\%) & Evnts & (fb) & (\%) & Evnts \\
\hline
HH($b\bar b \gamma\gamma$) & 0.089 & 6.2&16.6 & 0.545 & 5.04 & 82.4& 3.73 & 3.61 & 403.9\\
 \hline 
$b\bar b \gamma\gamma$ &294 &0.0045 & 40.1 &1085&0.0039 &126.4& 5037& 0.00275&415.4\\
$z(b\bar b) h(\gamma\gamma)$ & 0.109&1.48&4.86&0.278&1.41&11.8 &0.875& 1.57&41.2 \\
$b\bar b h(\gamma\gamma)$ & 2.23 &0.072&4.82 &9.84& 0.084&24.8&50.5& 0.099&150.5\\
$t\bar t h(\gamma\gamma)$ & 0.676 & 0.178&3.62 &4.76& 0.12&16.5& 37.3&0.11&124.2\\
\hline
Total B & -& -  & 53.4 &-& - & 179.5 &-& - & 731.3\\ \hline
S/$\sqrt{B}$ & - & - & 2.3 & - &-& 6.2 & - &-& 15.0 \\ \hline

\hline \hline
\end{tabular}
\end{table}

\begin{figure}[htpb]
\centerline{\psfig{file=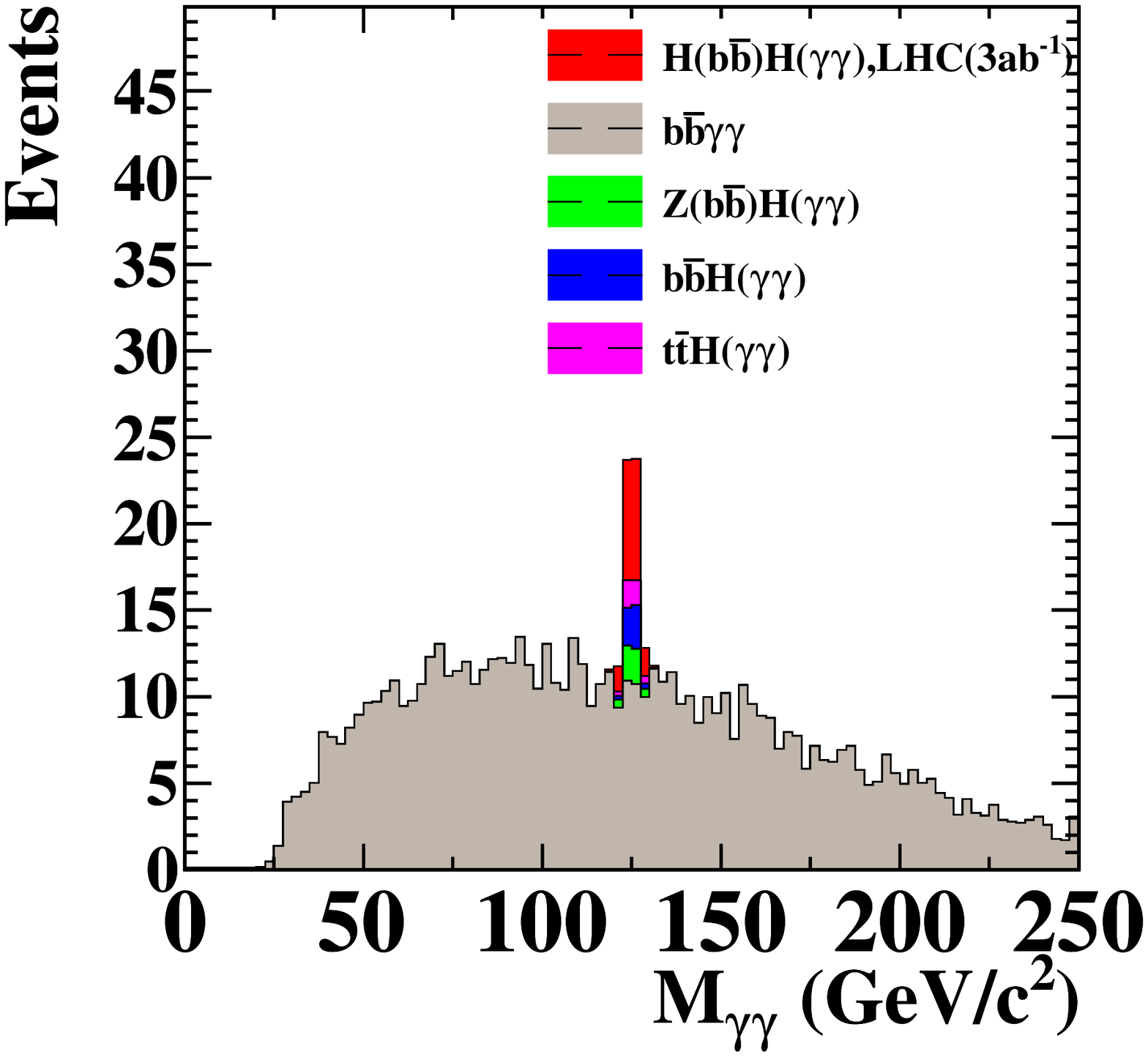,width=6.0cm}
\psfig{file=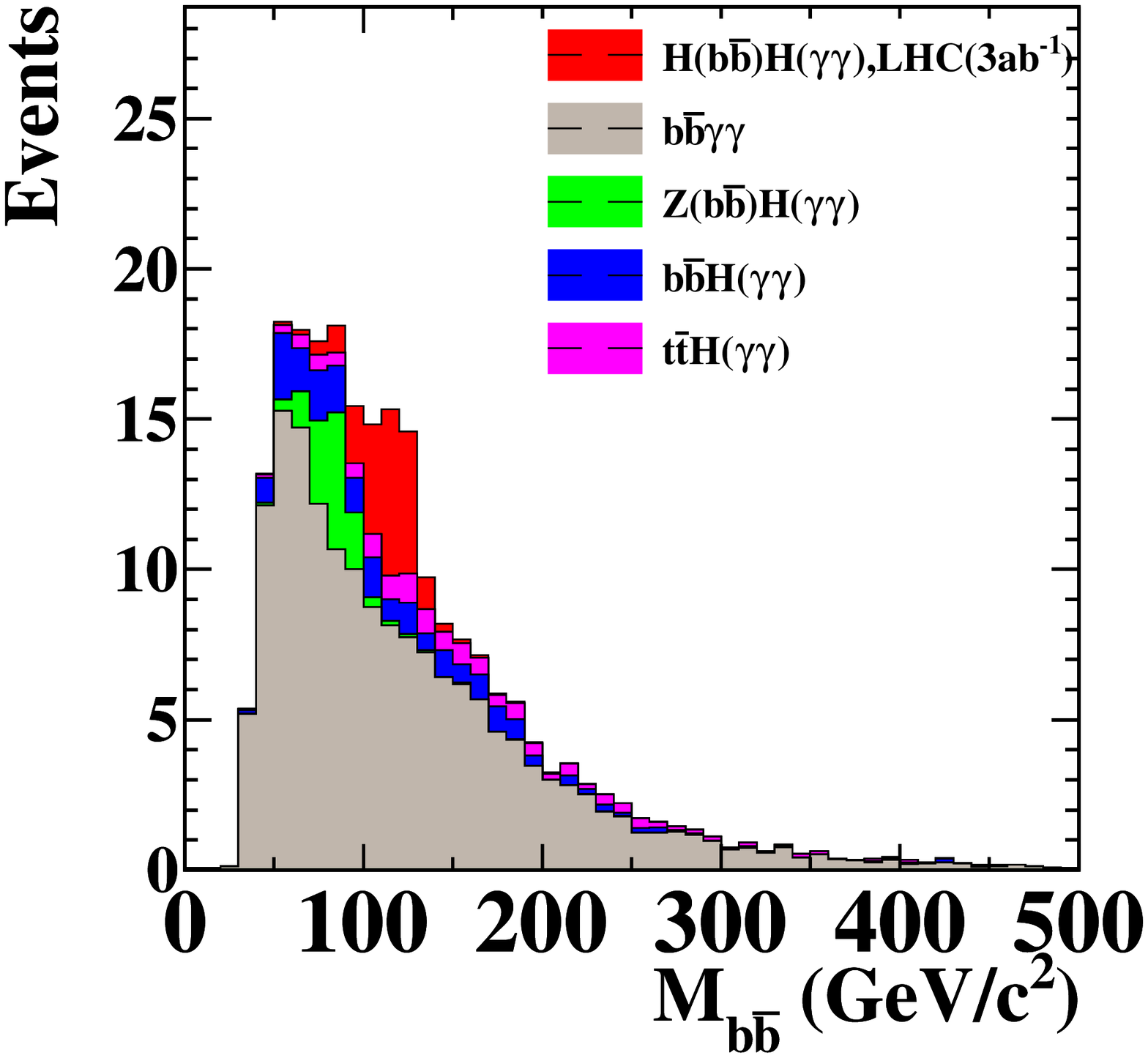,width=6.0cm}} %4.7cm
\vspace*{8pt}
\caption{The expected mass of two photons (left) or two $b$-jets (right) 
after requiring mass of two $b$-jets or two photons consistent 
with the Higgs mass at LHC with 3000 fb$^{-1}$ data.}
\label{fig:figlhc}
\end{figure}

\begin{figure}[htpb]
\centerline{\psfig{file=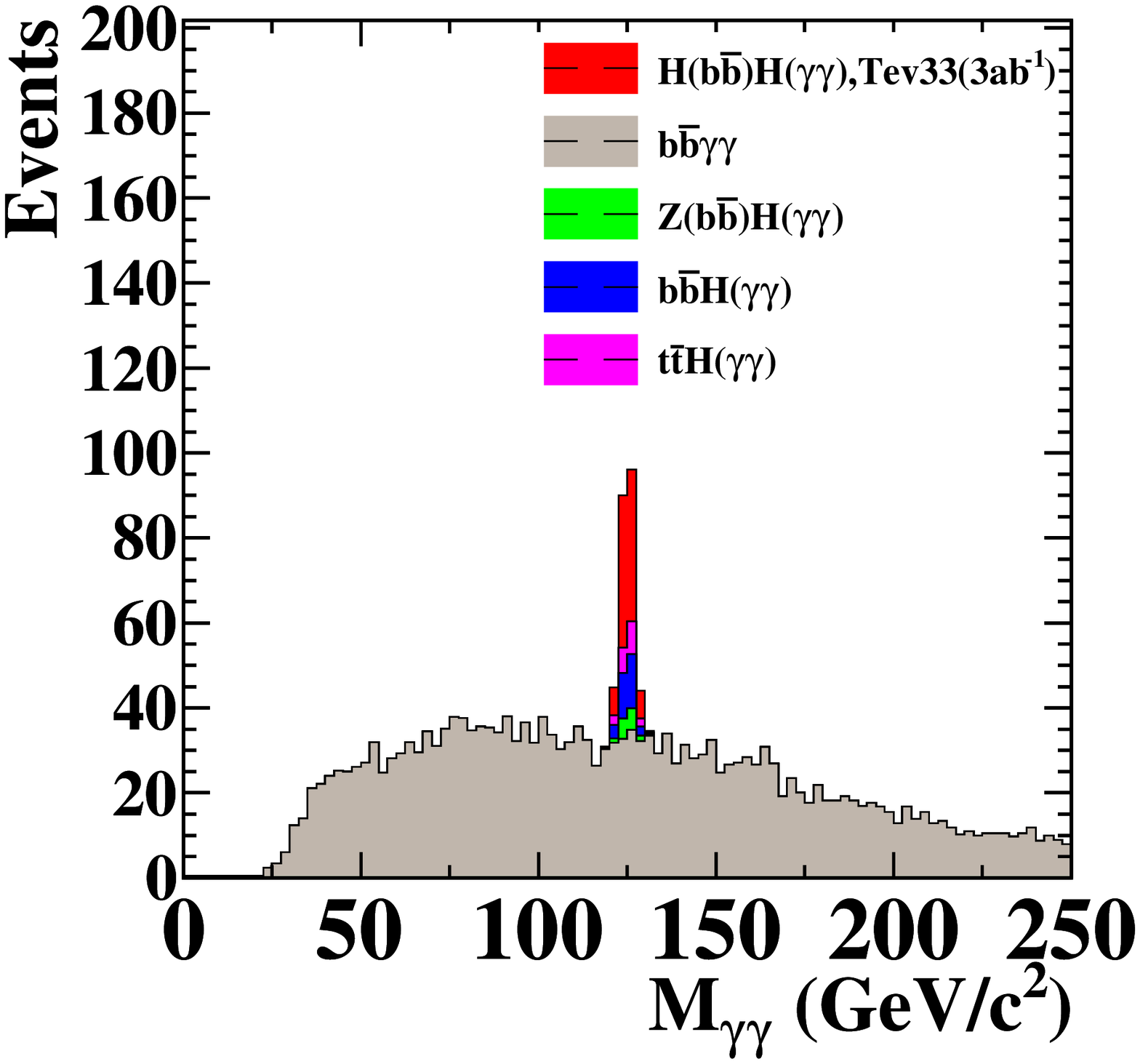,width=6.0cm}
\psfig{file=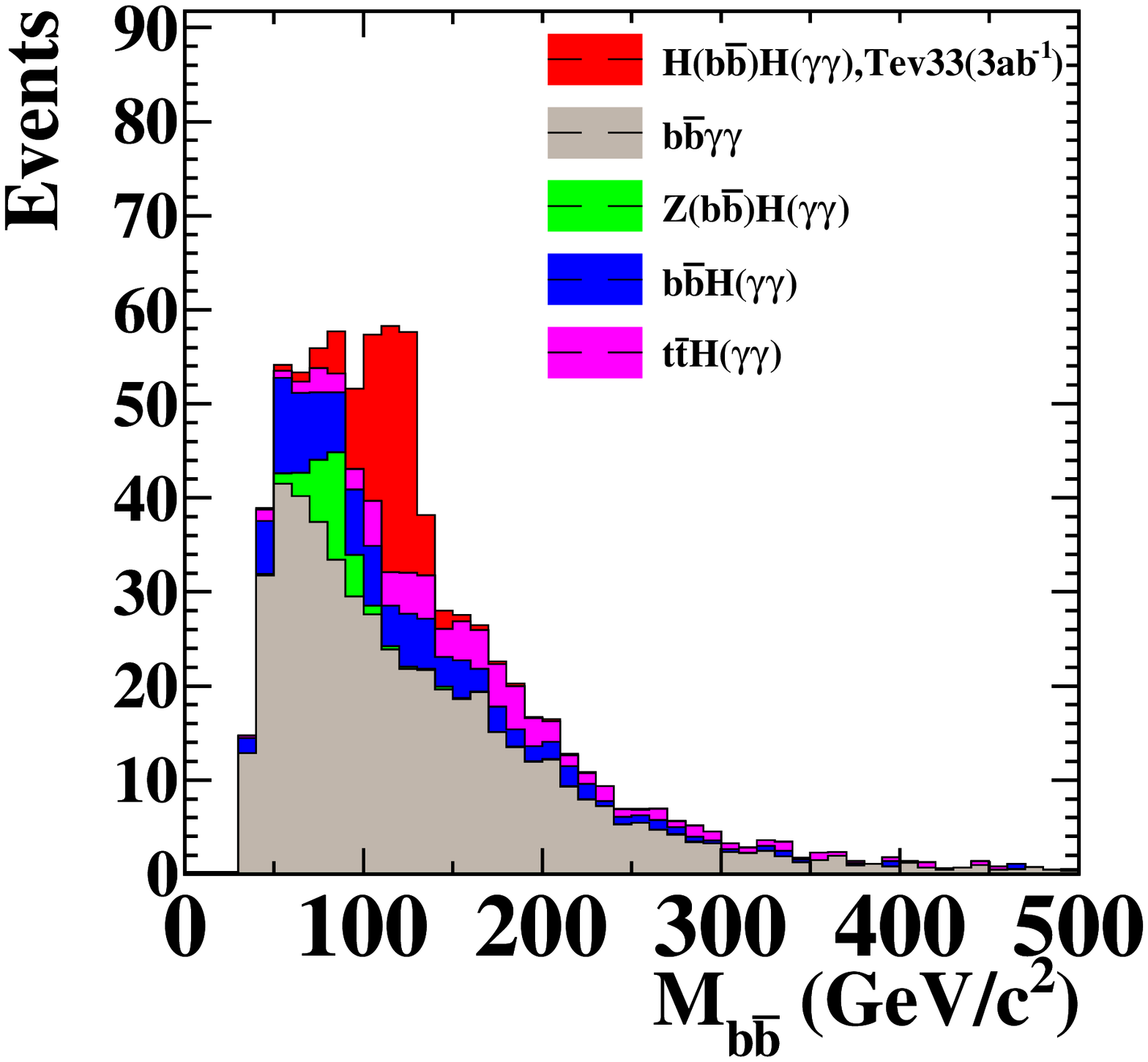,width=6.0cm}} %4.7cm
\vspace*{8pt}
\caption{The expected mass of two photons (left) or two $b$-jets (right) 
after requiring mass of two $b$-jets or two photons consistent 
with the Higgs mass at a Tev33 collider with 3000 fb$^{-1}$ data.}
\label{fig:figtev33}
\end{figure}

\begin{figure}[htpb]
\centerline{\psfig{file=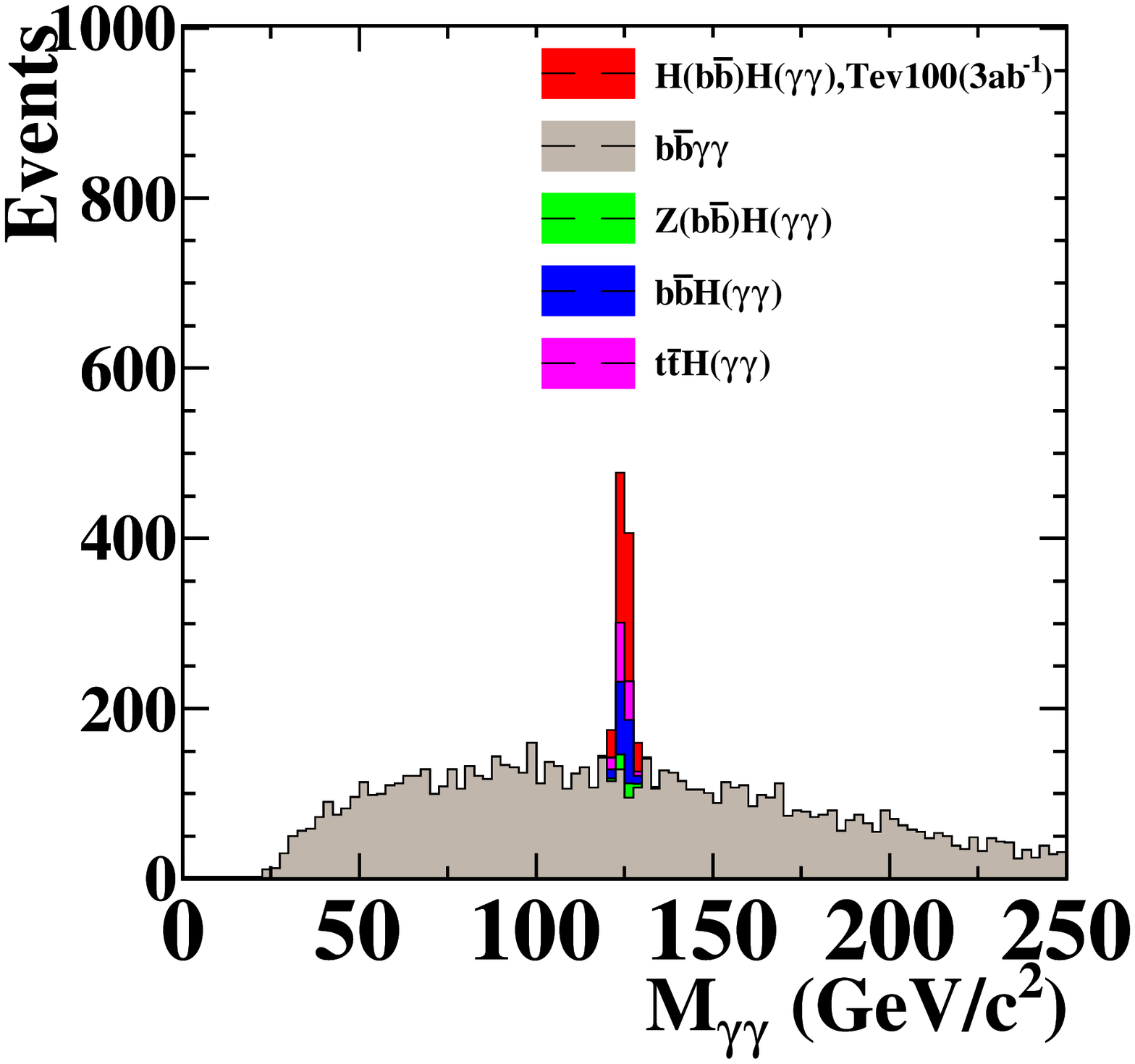,width=6.0cm}%6.7
\psfig{file=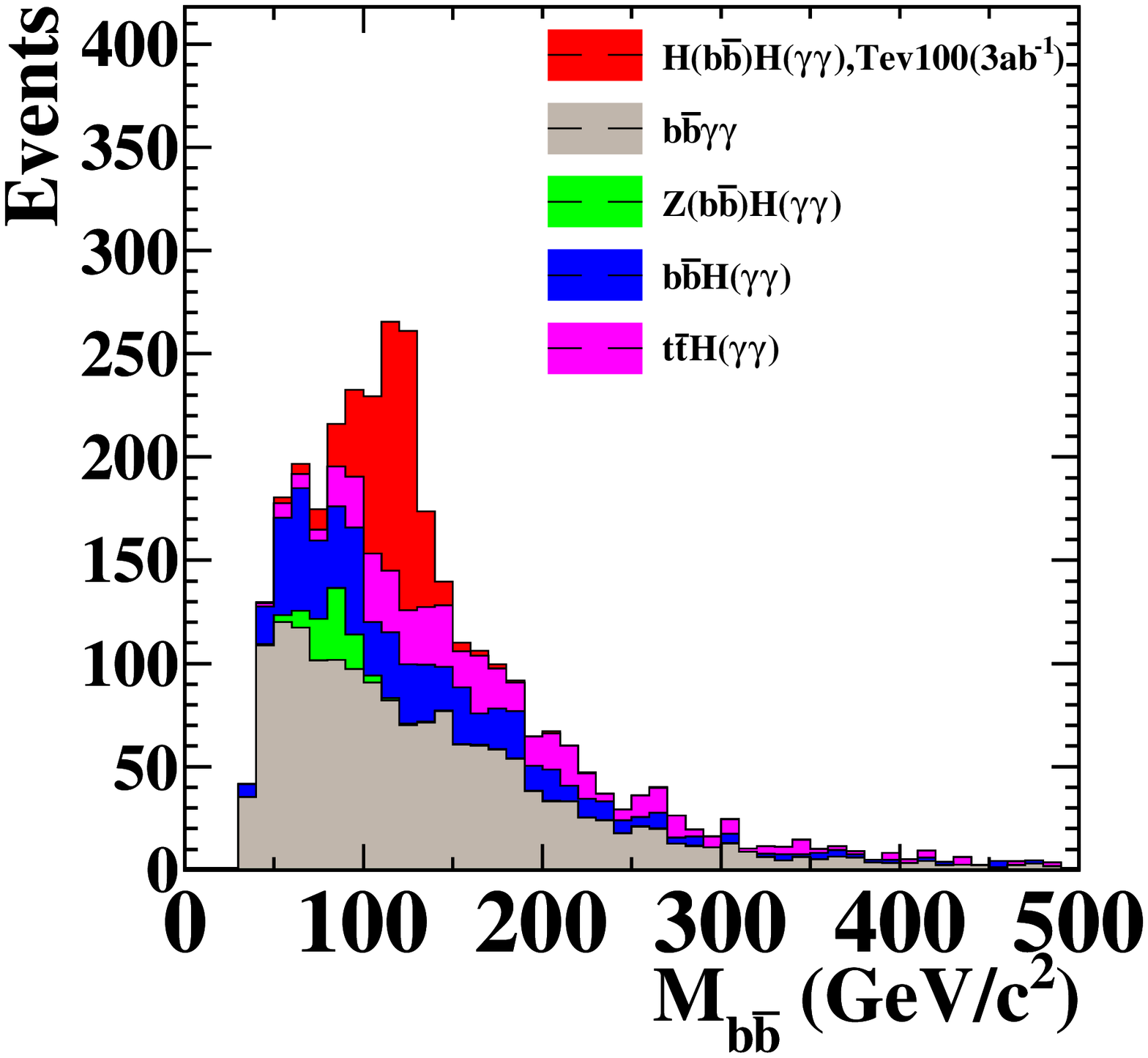,width=6.0cm}} %4.7cm
\vspace*{8pt}
\caption{The expected mass of two photons (left) or two $b$-jets (right) 
after requiring mass of two $b$-jets or two photons consistent 
with the Higgs mass at a Tev100 collider with 3000 fb$^{-1}$ data.}
\label{fig:figtev100}
\end{figure}

For the high luminosity running of LHC at 14 TeV,
it's possible to observe a statistical significance of 2.3 $\sigma$ signal with 
3000 fb$^{-1}$ data, which is consistent with the previous studies~\cite{atlashh}. For the 
higher energy colliders with $\sqrt{s}$=33, and 100 TeV, we would expect to observe 
a signal with a statistic significance of 6.2 and 15.0 $\sigma$ with 3000 
fb$^{-1}$ data, respectively. In Fig.~\ref{fig:figlhc}~-~\ref{fig:figtev100}, 
we show the projections 
of the final invariant mass of two photons or two $b$-jets after selecting 
$M_{b\bar b}$ or $M_{\gamma\gamma}$ for $\sqrt{s}=$14, 33, and 100 TeV colliders, 
respectively. After the $gg\rightarrow HH\rightarrow b\bar b\gamma\gamma$ 
signal is established, we would measure its production cross section and derive 
the Higgs self-coupling constants from the dependence of the 
$gg\rightarrow HH$ production cross section as a function of the Higgs 
self-coupling constants. Based on the estimation of 
$d(\sigma/\sigma_{SM})/d(\lambda/\lambda_{sm})\approx -0.8$ from 
Fig. 13 in ref.~\cite{baglio} and the significance of 
$HH\rightarrow bb\gamma\gamma$ signal, the Higgs self-coupling can be measured
to be a statistical accuracy of 50\%, 20\%, and 8\% with 3 ab$^{-1}$ data 
at the future colliders with $\sqrt{s}$=14, 33, and 100 TeV, respectively. 
However, it is worth to note that the event acceptance needs a correction for the 
dependence of Higgs self-coupling due to tight cuts used. 
In the future, we may have to loose some of selections while exploring 
kinematc distributions (shapes) that are most sensitive to the Higgs self-coupling to improve
the measurement beyond simple event counting. 

\section{Conclusion} 
We present a feasibility study of observing $HH\rightarrow b\bar b\gamma\gamma$
at the future hadron colliders with $\sqrt{s}=$14, 33, and 100 TeV. 
The measured cross section then can be used to 
constrain the Higgs self-coupling directly in the standard model. 
Any deviation could be a sign of new physics. The signal and background events 
are estimated using Delphes 3.0.10 fast Monte Carlo simulation based on the 
ATLAS detector capabilities. With 3 ab$^{-1}$ data, it would be possible to 
measure the Higgs self-coupling with a 50\%, 20\%, and 8\% statistical accuracy 
by observing $HH\rightarrow b\bar b\gamma\gamma$ at 
$\sqrt{s}=$14, 33, and 100 TeV colliders, respectively.  

\section{Acknowledgement}
We would like to thank C. Barrera, A. Nisati and N. Styles for their useful cross checks and valuable discussions.


\begin{thebibliography}{0}    %for 1 digit

%%journal paper

\bibitem{discovery}ATLAS Collab. (G. Aad {\it et al}.), {\it Phys. Lett. B\/} {\bf 716}, 1 (2012);\\
 CMS Collab. (S. Chatrchyan {\it et al}.), {\it Phys. Lett. B\/} {\bf 716}, 30 (2012).
 
\bibitem{higgs} F. Englert and R. Brout, {\it Phys. Rev. Lett.} 
{\bf 13}, 321 (1964); \\
P.W. Higgs, {\it Phys. Rev. Lett.} {\bf 13}, 508 (1964); \\
G.S. Guralnik, C.R. Hagen, and T.W.B. Kibble, {\it Phys. Rev. Lett.} 
{\bf 13}, 585 (1964). 

\bibitem{ilc} M. Battaglia, E. Boos, and W. Yao, Studying the Higgs Potential at the $e^+e^-$ Linear Collider, 
arXiv:hep-ph/011127.
\bibitem{baur04} U. Baur, T. Plehn, and D.Rainwater, {\it Phys. Rev. D.} {\bf 69}, 053004 (2004).
\bibitem{zurita} M.J. Dolan, C. Englert and M. Spannowsky, {\it JHEP} {\bf 10} 112 (2012);\\
  F. Goertz {\it et al.}, {\it JHEP} {\bf 06} 016 (2013);\\
  A. Papaefstathiou, L.L. Yang, and J. Zurita,{\it Phys. Rev. D.} {\bf 87} 011301(R) 2013.

\bibitem{baglio} J. Baglio {\it et al.}, The measurement of the Higgs self-coupling at the LHC:
theoretical status, arXiv:1212.5581.

\bibitem{atlashh} The ATLAS Collaboration, Studies of the ATLAS potential for Higgs self-coupling
measurements at a High Luminosity LHC, ATLAS-PHYS-PUB-2013-001.

\bibitem{NNLO} D.Y. Shao {\it et al.}, {\it JHEP} {\bf 07}, 169 (2013). 

\bibitem{delphes} S. Ovyn, X. Rouby and V. Lemaitre, 
DELPHES a framework for fast simulation of a generic collider experiment, arXiv:0903.2225.
  
\bibitem{hpair} E. El Kacimi and R. Lafaye, Simulation of neutral Higgs pair production 
processes in PYTHIA using HPAIR matrix elements, ATL-PHYS-2002-015. 

\bibitem{madgraph}J. Alwall {\it et al.}, {\it JHEP} {\bf 1106}, 128 (2011).

\bibitem{pythia8.0}T. Sjostrand, S. Mrenna, P. Skands, {\it Comput. Phys. Commun.} {\bf 178} 852 (2008). 

\bibitem{mlm}M.L. Mangano {\it et al.}, {\it JHEP} {\bf 07} 001 (2003).

\bibitem{cteq6l}K. Kovarik et al, CTEQ nuclear parton distribution functions, arXiv:1307.3454.  

\end{thebibliography}
\end{document}